\documentclass[conference]{IEEEtran}
\usepackage{textcomp}
\usepackage{pgfplots}
\pgfplotsset{width=8cm,compat=1.9}
\usepackage{hyperref}
\hypersetup{hidelinks} 
\usepackage{verbatim}
\usepackage{cite}
\usepackage{amsmath,amssymb,amsfonts}
\usepackage{algorithmic}
\usepackage{graphicx}
\usepackage{textcomp}
\usepackage{caption}
\usepackage{subcaption}
\usepackage{siunitx}
\usepackage[per-mode = symbol, free-standing-units,unit-optional-argument,space-before-unit,use-xspace]{siunitx}

\begin{document}

\title{Adaptive Attention-Based Model for 5G Radio-based Outdoor Localization}
\author{\IEEEauthorblockN{Ilayda Yaman\IEEEauthorrefmark{1}, Guoda Tian\IEEEauthorrefmark{3}, Dino Pjanić\IEEEauthorrefmark{3}, Fredrik Tufvesson\IEEEauthorrefmark{1}, Ove Edfors\IEEEauthorrefmark{1},  
Zhengya Zhang\IEEEauthorrefmark{2}, Liang Liu\IEEEauthorrefmark{1}}
\\
\IEEEauthorblockA{\IEEEauthorrefmark{1}Dept. of Electrical and Information Technology, Lund University, Sweden}
\IEEEauthorblockA{\IEEEauthorrefmark{3}Ericsson AB, Lund, Sweden}
\IEEEauthorblockA{\IEEEauthorrefmark{2}Department of Electrical Engineering and Computer Science, University of Michigan, Ann Arbor, USA}
ilayda.yaman@eit.lth.se, guoda.tian@ericsson.com, dino.pjanic@ericsson.com, fredrik.tufvesson@eit.lth.se, \\  ove.edfors@eit.lth.se, zhengya@umich.edu, liang.liu@eit.lth.se \\ 
}

\maketitle

 
\begin{abstract} 
Radio-based localization in dynamic environments, such as urban and vehicular settings, requires systems that efficiently adapt to varying signal conditions and environmental changes. Factors like multipath interference and obstructions introduce different levels of complexity that affect the accuracy of the localization. Although generalized models offer broad applicability, they often struggle to capture the nuances of specific environments, leading to suboptimal performance in real-world deployments. In contrast, specialized models can be tailored to particular conditions, enabling more precise localization by effectively handling domain-specific variations, which also results in reduced execution time and smaller model size. However, deploying multiple specialized models requires an efficient mechanism to select the most appropriate one for a given scenario. In this work, we develop an adaptive localization framework that combines shallow attention-based models with a router/switching mechanism based on a single-layer perceptron. This enables seamless transitions between specialized localization models optimized for different conditions, balancing accuracy and computational complexity. We design three low-complex models tailored for distinct scenarios, and a router that dynamically selects the most suitable model based on real-time input characteristics. The proposed framework is validated using real-world vehicle localization data collected from a massive MIMO base station and compared to more general models.
\end{abstract}


\section{Introduction}

Reliable and accurate location information with low latency is essential to ensure efficient and reliable workflows in various tasks, including vehicle navigation, intelligent traffic management, and autonomous driving. Localization in urban environments presents significant challenges due to multipath propagation, signal blockages, and dynamic environmental conditions. Traditional Global Navigation Satellite System (GNSS)-based positioning often suffers from poor accuracy in dense urban areas, where buildings and other structures obstruct satellite signals. A robust and continuously available localization solution is needed that works seamlessly across different environments and platforms~\cite{mixed_los_nlos}. For example, in~\cite{adverse_outdoor}, the authors implement random forest and gradient boosting algorithms to effectively use multipath information to improve outdoor location. When tested on real-world data, machine learning (ML) models achieve a mean localization error of approximately $100$\,m in an area of $580,000\textrm{\,m}^{2}$.  
 
A key challenge in radio-based localization is designing computationally efficient models capable of adapting to dynamic signal conditions. Several works in the literature have considered using massive multiple-input-multiple-output (MIMO) technologies together with ML methods to achieve robust, low-latency localization in diverse scenarios~\cite{tian2024attention, vtc2020, mimo_outdoor}. A widely adopted approach is fingerprinting, where channel state information or channel impulse response (CIR) measurements serve as unique signatures of specific locations within an environment. By collecting and storing these measurements in a database, ML models can be trained to map real-time channel responses to their corresponding spatial coordinates $(x, y)$. Deep learning-based fingerprinting methods have demonstrated high localization accuracy by taking advantage of the spatial and temporal characteristics of the wireless channel~\cite{icl_gnss2022, tian2023deep, urban}.

Specialized fine-tuned models can achieve higher accuracy with smaller model sizes and less computational complexity compared to generalized models. 
To accommodate multiple conditions, general models are more complex, use more parameters, and have higher test times and computational costs. However, to fully leverage the advantages of specialized models across diverse environments, a router is needed to dynamically determine which model is best suited for a given scenario. By analyzing input characteristics and environmental conditions, the router selects the most appropriate specialized model, ensuring that only the necessary parameters are activated at each time step. Due to the continuous nature of the application, switching between different models would happen rarely, and the weights would be reused extensively. Thus, computational overhead is reduced and high localization accuracy is maintained under different conditions. By integrating adaptive models, localization systems dynamically adjust to varying urban scenarios, ensuring reliable and efficient positioning for applications such as autonomous driving, intelligent transportation, and urban navigation. 

\begin{figure*}[tb!]
  \centering
  \includegraphics[width=0.8\linewidth]{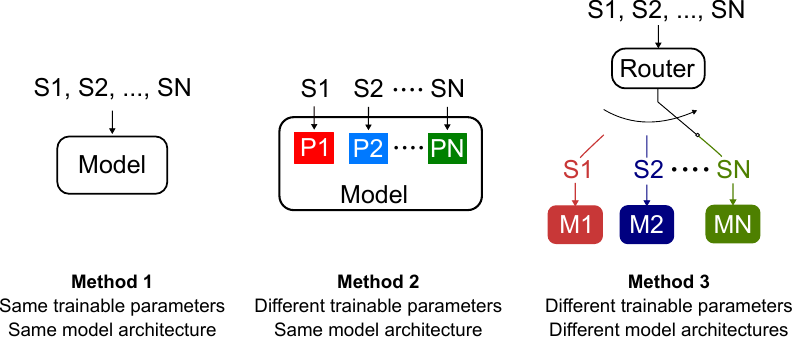}
  \caption{Overview of the adaptive model, where S, N, P, and M represent the subset of data, the total number of subsets, the trainable parameters (weights and biases), and the model, respectively}
  \label{fig:adaptive_model}
\end{figure*}

Fig.~\ref{fig:adaptive_model} visualizes three methods, with Method~1 being the most general and Method~3 the most specialized. 
Method~1 uses the same trainable parameters and architecture in all data subsets.
Method~2 consists of $N$ sets of trainable parameters that are loaded into the same model depending on the selected subset of data. Switching between trainable parameters is done manually. 
The proposed adaptive model, Method~3, includes specialized trainable parameters and a model architecture with run-time adaptation between different data subsets. The router in Method~3 determines which scenario the given input belongs to and selects the appropriate model accordingly, ensuring that only a subset of all parameters is active.

By tuning the model architecture parameters, the specialized models provide better localization accuracy than the other methods. The number of active trainable parameters and the training and testing times are also reduced. The main contributions of the paper are: 


\begin{itemize}
    \item We establish specialized attention-based models to minimize computational complexity, localization error, and model size for different scenarios. We then compare the results with more general models.  
    \item We develop an adaptive localization framework capable of dynamically adjusting to environmental changes and signal propagation conditions. Measurement data from a massive MIMO base station (BS) and a moving user equipment (UE) on a vehicle is used for verification.   
\end{itemize}


\section{Fundamentals of Attention Models and Experimental Validation} 

This section provides an overview of the key components of attention-based models and describes the measurement campaign conducted to validate the proposed model.

\subsection{Building Blocks of Attention-based Models}

\begin{figure}[tb!]
  \centering
  \includegraphics[width=1.0\linewidth]{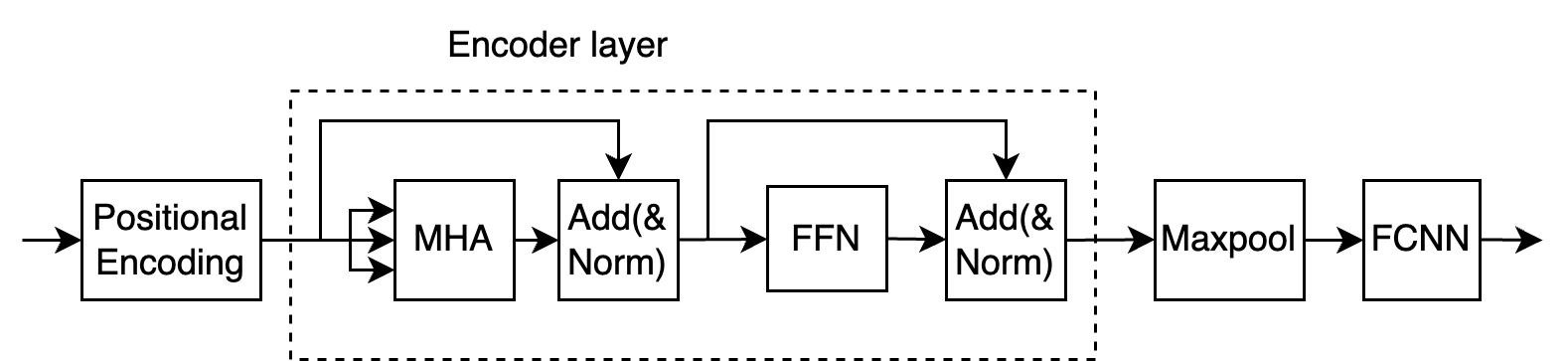}
  \caption{The attention-based algorithm pipeline with one encoder layer and pooling layer added.}
  \label{fig:detailed_diagram}
\end{figure}

The architecture of the shallow encoder-only attention-based model is shown in Fig.~\ref{fig:detailed_diagram}. The model consists of positional encoding, multihead attention, and feedforward layers, each followed by residual connections and optional layer normalization. Additionally, a pooling layer is applied before the fully connected neural network (FCNN) to reduce the size of the model. 

\subsubsection{Positional Encoding} is used to incorporate position information into the model, using sinusoidal functions to generate continuous and differentiable positional vectors. This enables the model to capture relative positional relationships effectively. The model is tested with and without positional encoding, showing a significant drop in accuracy when it is removed.

\subsubsection{Multi-Head Attention (MHA)}
is the fundamental mechanism in attention-based architectures, enabling efficient parallel processing of input data~\cite{attentionisallyouneed}. It is based on Scaled Dot-Product Attention, which computes attention scores based on query, key, and value matrices derived from the input tensor \( \mathbf{X} \in \mathbb{R}^{d_k \times d_{\text{model}}} \) at a given timestep

\begin{equation}
    \mathbf{Q} = \mathbf{X}\hspace{1pt}\mathbf{W}_q, \hspace{2pt}
    \mathbf{K} = \mathbf{X}\hspace{1pt}\mathbf{W}_k, \hspace{2pt}
    \mathbf{V} = \mathbf{X}\hspace{1pt}\mathbf{W}_v.
\end{equation}
The attention mechanism is computed as

\begin{equation}
    \text{Attention}(\mathbf{Q}, \mathbf{K}, \mathbf{V}) = \text{softmax} \left( \frac{\mathbf{Q} \mathbf{K}^T}{\sqrt{d_k}} \right) \mathbf{V},
\end{equation}
where \( d_k \) is the dimensionality of the key vectors, and the scaling factor \( \sqrt{d_k} \) prevents large softmax values that could hinder training. The MHA extends this by applying multiple attention mechanisms in parallel, each with different learned projections

\begin{equation}
    \text{head}_i = \text{Attention}(\mathbf{Q} \mathbf{W}_i^Q, \mathbf{K} \mathbf{W}_i^K, \mathbf{V} \mathbf{W}_i^V),
\end{equation}
where \( \mathbf{W}_i^Q, \mathbf{W}_i^K, \mathbf{W}_i^V \) are learned projection matrices for the \( i \)-th attention head. The outputs of all heads are concatenated and projected to obtain the final representation:
\begin{equation}
    \text{MHA}(\mathbf{Q}, \mathbf{K}, \mathbf{V}) = \text{Concat}(\text{head}_1, \dots, \text{head}_h) \mathbf{W}^O,
\end{equation}
where \( h \) is the number of attention heads, and \( \mathbf{W}^O \) is a learned output projection matrix. By allowing the model to attend to different representation subspaces simultaneously, MHA enhances its ability to capture complex dependencies, e.g. in challenging environments where noise and signal multipath effects complicate traditional methods.


\subsubsection{Layer Normalization (LN)} normalizes activations across the feature dimension for each input independently, making it effective for stabilizing deep networks.

\subsubsection{Position-wise Feed-Forward Networks (FFN)} applies two linear transformations with an activation function in between. Each token in the sequence is processed independently through a position-wise transformation. The dimensionality of the inner layer is represented as $d_{\text{ff}}$. The activation function is the rectified linear unit (ReLU). 

\subsubsection{Pooling layer} is used to reduce the model size and computational complexity while retaining key feature representations. A one-dimensional max pooling layer is introduced between the encoder and the fully connected network. Max pooling is preferred over alternatives such as mean pooling due to its hardware efficiency, as it requires only comparison operations rather than arithmetic computations. 
Padding can be added to adjust alignment. This approach effectively reduces the number of parameters and computations in subsequent layers, resulting in a more compact and efficient model.

\subsubsection{Dropout} is a regularization technique that helps prevent overfitting by randomly deactivating neurons during training. The dropout is added after the positional encoding, softmax, eq. 4 and pooling, and before the residual connections seen in Fig. 2. Different dropout rates were explored (0, 0.05, 0.1, 0.2, 0,3 and 0.5) and 0.05 was chosen based on the accuracy and training/validation loss of the model. 

\subsection{Experiments}

The measurement campaign is based on a commercial self-contained massive MIMO BS and a single mobile user in a vehicle. The vehicle is equipped with a UE and a GNSS receiver, which is used for ground truth. The BS operates at $3.85$\,GHz center frequency and $100$\,MHz bandwidth. The uplink (UL) sounding reference signal (SRS) channel estimates span 273 Physical Resource Blocks (PRBs) across the full bandwidth, with each snapshot capturing all 64 beams every 20 ms. To reduce complexity during data collection, adjacent PRBs are averaged and downsampled into 137 subgroups (SGs), which are then further reduced to 46 PRBSGs (subcarriers) by interleaving every third SG. The SRS measurements represent the angular delay spectrum of the radio channel in the beam space after the FFT transformation from the antenna space.

The BS is equipped with $32$ vertically and $32$ horizontally polarized antenna ports, while the UE includes $4$ antenna ports. The elements of the antenna array are used to form $64$ beams in both the UL and the downlink. The UL SRS data sounded from $2$ paired UE antenna ports at the time are recorded across $46$ subcarriers at the BS. The $32$ horizontal and vertical beam space matrices of the channel transfer function (CTF) from the first and second antenna pairs at time $t$ and across $46$ subcarriers are denoted as $\mathbf{H}_{\text{V1,t}}$, $\mathbf{H}_{\text{H1,t}}$, $\mathbf{H}_{\text{V2,t}}$ and $\mathbf{H}_{\text{H2,t}}$, respectively. 
The combined channel matrix corresponds to $\mathbf{H}_t \in \mathbb{C}^{128 \times 46} = \left[\mathbf{H}_{\text{H1},t}^T,\mathbf{H}_{\text{V1},t}^T, \mathbf{H}_{\text{H2},t}^T,\mathbf{H}_{\text{V2},t}^T\right]^T$ for each channel snapshot. 

In the pre-processing step, invalid data is filtered out, and a 46-point Hann window is applied across the rows of the CTF to suppress sidelobes. Subsequently, an IFFT is performed along the x-axis (i.e., across rows) to transform CTF into a CIR beam matrix that captures angular information. The amplitude values of the CIR beam matrix are then used as input to the attention-based model.

\begin{figure}[tb!]
  \centering
  \includegraphics[width=1.0\linewidth]{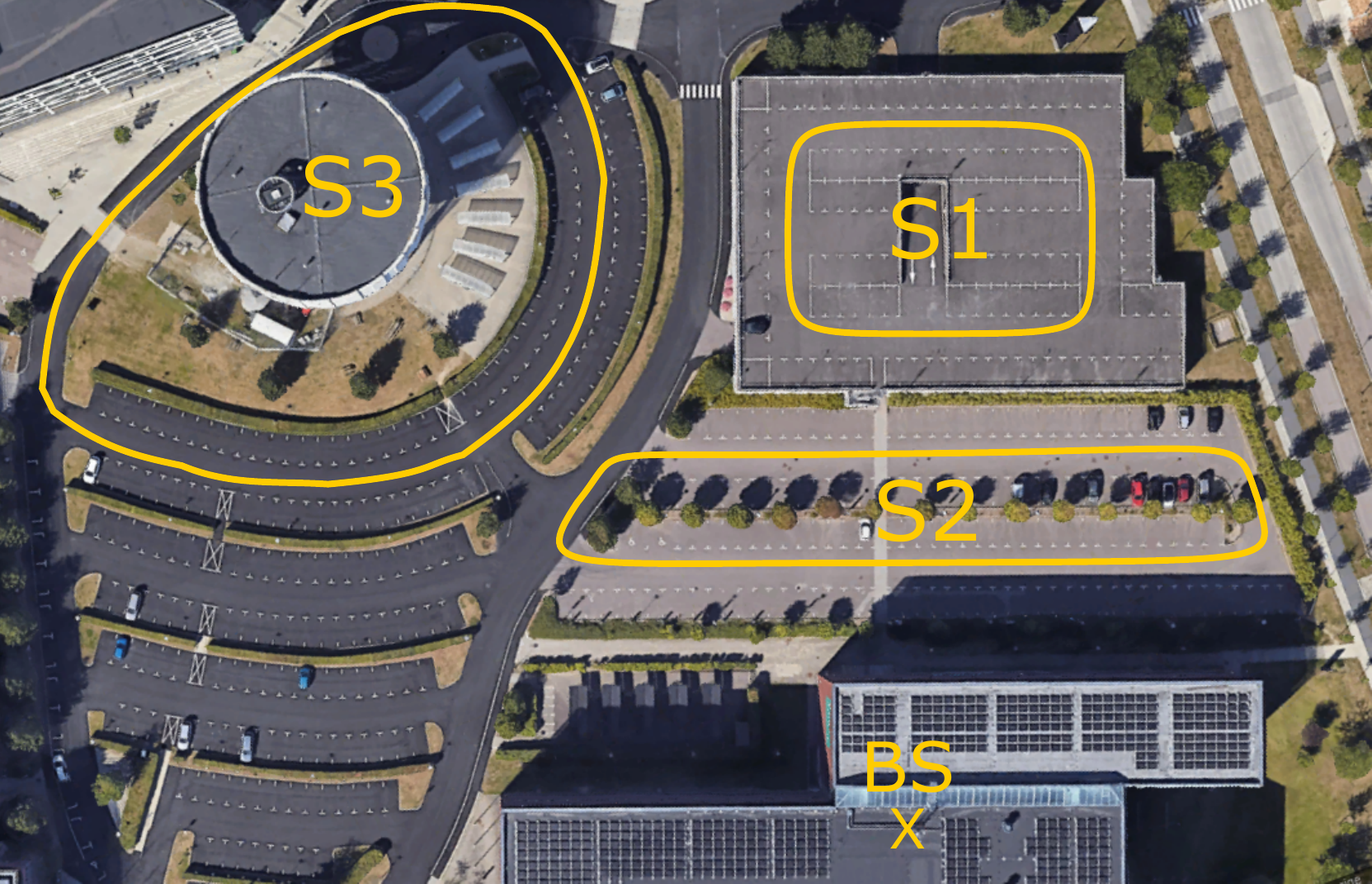}
  \caption{Bird-eye view of the measurement environment and trajectories labeled S1, S2, and S3.}
  \label{fig:measurement_env}
\end{figure}

The UE is mounted on a vehicle following three predefined trajectories, forming three distinct scenarios, shown by yellow lines in Fig.~\ref{fig:measurement_env}. The BS location is marked by a yellow cross. The UE antenna maintains a consistent orientation as it is fixed to the vehicle roof and follows the same orientation for each lap. The average speed of the vehicle is $15 \text{\,km} /\text{\,h}$ and the BS is mounted on a 20-meter-high building. The vehicle follows the given trajectories in the same direction for five laps.
For more detailed information on the measurement campaign and pre-processing of the measurement data, see~\cite{tian2024attention}.

In the first scenario (S1), the vehicle drives on the roof of a garage, approximately 10 meters above ground level, following a trajectory that predominantly captures line-of-sight (LoS) conditions. The second scenario (S2) takes place at ground level beneath the base station (BS), where the vehicle is mostly not in direct LoS with the BS. The third scenario (S3), also at ground level, features intermittent LoS along the trajectory, with obstructions introduced by the tower visible in Fig. 3. These three scenarios constitute our data subsets: S1, S2, and S3. From the BS’s perspective, users typically follow predefined urban mobility patterns. However, we acknowledge the repetitive nature of the selected UE trajectory and the fixed antenna orientation toward the BS, as the UE was securely mounted on the vehicle roof, thereby reducing variability in the radio channel.

\subsection{Evaluation Method}

Of the five laps, the first four laps are used for training and validation, whereas the fifth lap of each scenario is reserved for testing. For the generalized model, the training/validation datasets of different scenarios are randomly mixed to create a dataset that consists of different scenarios, whereas the specialized models are only trained on the specific trajectories.

The accuracy of the localization algorithms is calculated using the mean of the Euclidean distance between the estimated locations and the ground truth labels. Mean Euclidean Error (MEE) is defined as
\begin{equation}
    \text{MEE} = \frac{1}{N} \sum_{i=1}^{N} ||\mathbf{\hat{y}}_i - \mathbf{y}_i||_F
\end{equation}
where $N$, $\mathbf{\hat{y}}_i$ and $\mathbf{y}_i$ represent the number of samples, predicted $2$D locations and the ground truth labels, respectively. $||\cdot||_F$ is the Frobenius norm. 

\section{Proposed method}

The model architecture selection of specialized models and the router that is used to dynamically switch between different scenarios are described in this section. The hyperparameter configuration of the model is summarized in Table \ref{tab:parameters}. For more information on the selection of the hyperparameters, please see \cite{tian2024attention}. 
The input of the model is the CIR beam matrix for each single snapshot described in the previous section. 

\begin{table}[htbp!]
\caption{Hyperparameters of the model.}
\begin{center}
\begin{tabular}{|l|c|}
     \hline
     \textbf{Parameter} & \textbf{Value}  \\ \hline
     Epochs & $200$  \\\hline
     Batch Size (b) & $64$ \\\hline
     $ d_{\text{model}} $ & $46$ \\\hline
     $ d_k $ & $128$ \\\hline
     $d_{\text{ff}}$ & $64$ \\\hline   
     Learning Rate & $0.0006$  \\\hline
     Number of heads & $2$ \\\hline
     Dropout Rate & $0.05$  \\\hline  
     Loss function & MSE \\\hline
  \end{tabular}
  \label{tab:parameters}
  \end{center}
\end{table}

\subsection{Specialized models} 

\begin{figure}[tb!]
  \centering
  \includegraphics[width=1.0\linewidth]{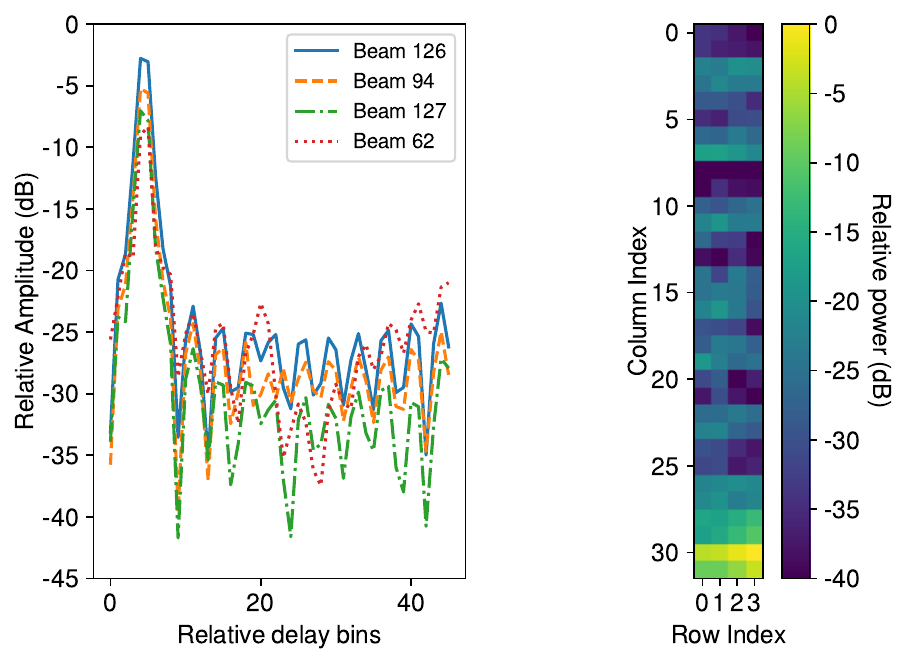}
  \caption{Power delay profile of the $4$ dominant beams and relative power of all $128$ beams in a LoS scenario.}
  \label{fig:LoS_Figure}
\end{figure}

The complexity of the model design is closely linked to the characteristics of the channel state information, which is highly influenced by multipath propagation. In an non LoS (NLoS) environment, the CIR and angular characteristics would differ significantly from a scenario where there is a dominant LoS component. In LoS environments, the energy is typically concentrated in the LoS path; therefore, the power delay profile changes slowly with UE movement. 
For example, in Fig.~\ref{fig:LoS_Figure}, the power is concentrated in the few early delay bins. The relative power of each beam is calculated by summarizing the power among all delay bins.  

\begin{figure}[tb!]
  \centering
  \includegraphics[width=1.0\linewidth]{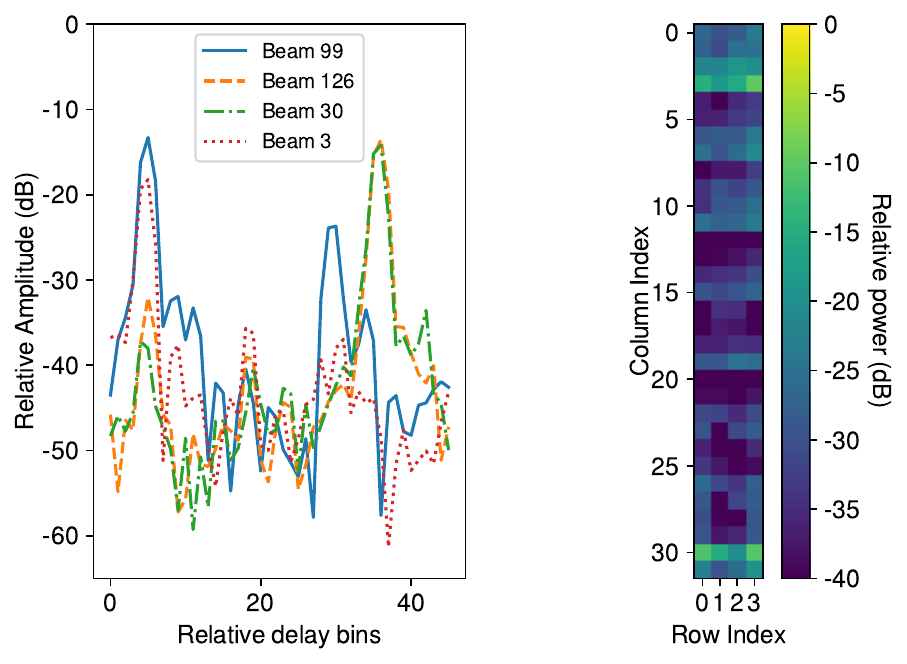}
  \caption{Power delay profile of the $4$ dominant beams and relative power of all $128$ beams in an NLoS scenario.}
  \label{fig:NLoS_Figure}
\end{figure}

In rich scattering environments (e.g., dense urban areas, indoor clutter), multipath reflections originate from numerous uncorrelated scatterers, resulting in less structured and more random components. Figure~\ref{fig:NLoS_Figure} illustrates this with a power delay profile and the corresponding beam matrix in an urban NLoS scenario. Unlike LoS conditions, the power delay profile in urban NLoS environments changes rapidly with UE position, motivating the design of specialized models to better capture such dynamics. Deeper models are more effective at modeling complex multipath interactions in these scenarios, whereas simpler models suffice under LoS conditions.

The following model architecture parameters are explored to select specialized models for different scenarios with different complexities: the effect of increasing the number of encoder layers from $1$ to $5$, removing/adding layer normalization, and adding a pooling layer before the FCNN layer. The pooling layer downsamples the input to the FCNN layer by selecting the maximum value in non-overlapping segments of size $4$. Padding is added to adjust the alignment. The output dimension becomes $(b, d_k, (d_{\text{model}}+2)/4)$, resulting in a reduced number of trainable parameters. By exploring these parameters and comparing them to Method~1, the optimal specialized models are selected that improve the accuracy while keeping the size of the model similar for each scenario. 

The results are shown in Table \ref{tab:results_models13}, comparing the generalized (Method~1) and specialized (Method~3) models. The highest localization accuracy for Method~1 is achieved with 3 encoder layers, resulting in MEE of $(0.47\text{m}, 0.83\text{m}, 0.81\text{m})$ for S1, S2, and S3, respectively. The best results are achieved when layer normalization is added for S3 and removed for S1 and S2. For Model~3, a lower MEE for S1 can be achieved using $1$ encoder layer, with or without the addition of pooling and layer normalization. For S2 and S3, $2$ encoder layers are needed to achieve better performance with specialized models. Although adding more encoder layers can further reduce MEE, computational complexity and model size increase accordingly.


\begin{table}[htbp!]
\caption{Localization accuracy comparison between Method 1 and Method 3 across various model architectures (EL = Encoder Layers, LN = Layer Normalization, MP = MaxPool). Bold values indicate architectures achieving optimal balance between Mean Euclidean Error (MEE), computational complexity, and parameter count.} 
\begin{center}
\begin{tabular}{|c|c|c|ccc|ccc|}
     \hline
      &  & & \multicolumn{3}{c|}{MEE (m), Generalized} & \multicolumn{3}{c|}{MEE (m), Specialized} \\ \hline  
      \textbf{EL} &  \textbf{LN} & \textbf{MP} & \textbf{S1}  & \textbf{S2}  & \textbf{S3} & \textbf{S1}  & \textbf{S2}  & \textbf{S3} \\ \hline
      1 & - & - & 0.85 & 1.14 & 1.11 & 0.41 & 0.96 & 0.99  \\\hline
      2 & - & - & 0.53 & 0.95 & 1.00 & 0.40 & 0.80 & 0.93  \\\hline
      3 & - & - & \textbf{0.47} & \textbf{0.83} & 0.86 & 0.35 & 0.81 & 0.83  \\\hline
      4 & - & - & 0.51 & 1.02 & 0.86  & 0.33 & 0.80 & 0.60\\\hline
      5 & - & - & 1.47 & 3.10 & 2.28 & 0.49 & 6.16 & 0.59 \\\hline\hline
      1 & + & - & 0.69 & 1.18 & 0.97 & 0.45 & 1.06 & 0.97\\\hline
      2 & + & - & 0.54 & 1.00 & 1.05 & 0.37 & 0.88 & 0.74\\\hline
      3 & + & - & 0.55 & 0.86 & \textbf{0.81} & 0.37 & 0.71& 0.77 \\\hline
      4 & + & - & 0.60 & 0.98 & 0.76 & 0.34 & 0.80 & 0.65 \\\hline
      5 & + & - & 0.68 & 1.08 & 0.97 & 1.60 & 5.76 & 1.01\\\hline\hline
      1 & - & + & 0.93 & 1.80 & 1.62 & \textbf{0.40} & 1.10 & 1.26\\\hline
      2 & - & + & 0.64 & 1.14 & 1.24 & 0.38 & \textbf{0.78} & \textbf{0.77}\\\hline
      3 & - & + & 0.57 & 1.14 & 1.11 & 0.38 & 0.81 & 0.81\\\hline
      4 & - & + & 0.88 & 1.30 & 1.13 & 0.40 & 0.93 & 1.55\\\hline
      5 & - & + & 2.59 & 4.39 & 3.91 & 9.68 & 0.79 & 2.62\\\hline\hline
      1 & + & + & 1.13 & 1.57 & 1.84 & 0.61 & 1.10 & 1.25\\\hline
      2 & + & + & 0.76 & 1.20 & 1.03 & 0.37 & 0.89 & 0.75 \\\hline
      3 & + & + & 0.70 & 1.08 & 1.10 & 0.38 & 0.92 & 0.82\\\hline
      4 & + & + & 0.75 & 1.15 & 0.88 & 0.33 & 0.99 & 0.97\\\hline
      5 & + & + & 1.13 & 1.93 & 1.74 & 0.86 & 7.84 & 3.00\\\hline
  \end{tabular}
  \label{tab:results_models13}
  \end{center}
\end{table}


\subsection{Router} In the proposed model, the router determines the type of scenario based on the input, a task known as classification. ML classifiers such as the Single-Layer Perceptron (SLP), Multilayer Perceptron (MLP), and Convolutional Neural Networks (CNN) are widely used for classification tasks. Among SLP, MLP, and CNN, the SLP is the simplest and least computationally expensive, making it suitable for linearly separable problems. It consists of an input layer directly connected to an output layer, with no hidden layers in between. The SLP performs a linear transformation of the input and applies an activation function to produce the output, and defined as 
\begin{equation}
    \mathbf{y} = \sigma(\mathbf{x} W + b)
\end{equation}
where \( \mathbf{x} \in \mathbb{R}^n \) is the input vector, \( W \) is the weight vector, \( b \) is the bias term, \( \sigma(\cdot) \) is the activation function (typically sigmoid or softmax), and \( \mathbf{y} \) is the predicted output. The training and test data split used for the generalized attention-based model is also applied in this section. This strategy ensures that the training data were constructed by randomly sampling from all environments, ensuring that each batch contained a diverse set of propagation conditions. 

When the router is implemented as an MLP or CNN, it achieves a test accuracy of $100\%$. To investigate the effects of overfitting, the dropout is increased to $0.7$, and a weight decay of $1 \times 10^{-5}$ is introduced. As a result, the test accuracy decreases to $99\%$. These findings indicate that the router model can be further reduced in complexity, potentially to an SLP.  

With the complete input matrix, the training and validation losses of the SLP model converge to zero, and the test accuracy remains at $100\%$.
To explore further model compression, trials are conducted using a smaller model that trains on the beam power matrix of one relative delay bin. The corresponding input is fed into a compact model with $387$ parameters, achieving a test accuracy of $98.87\%$.
Additional trials in different relative delay bins reveal that the minimum test accuracy observed is $97.76\%$. Given that there are $46$ relative delay bins in total, this shows that using a subset of input features can still produce high classification accuracy.  


\section{Results} 

\begin{figure}[tb!]
  \centering
  \includegraphics[width=1.0\linewidth]{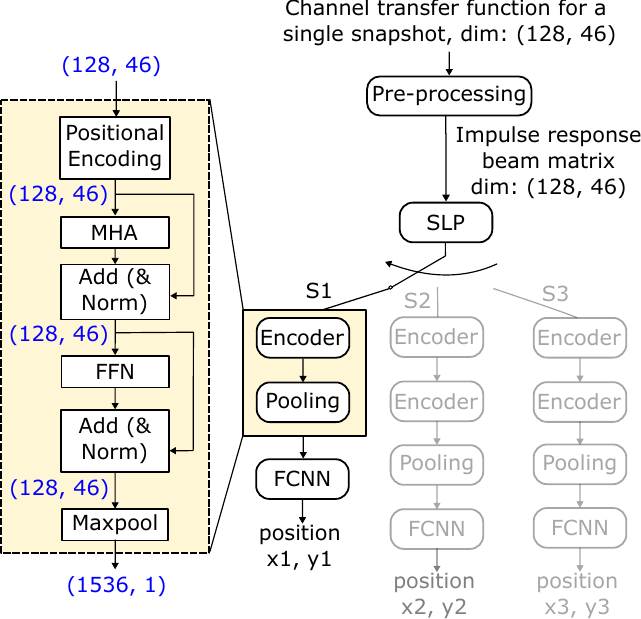}
  \caption{The final block diagram of the proposed method after selecting the specialized models and the SLP router. The positional encoding is integrated into the first encoder layer for simplicity in the diagram. Input data dimensions at each layer are indicated in blue. }
  \label{fig:high_level_diagram}
\end{figure}

The final proposed architecture is depicted in Fig.~\ref{fig:high_level_diagram}. The diagram illustrates the active model with black lines, while the inactive models appear in gray. The sub-blocks of the "Encoder" layer in Fig.~\ref{fig:high_level_diagram} are given in Fig.~\ref{fig:detailed_diagram}. 

Table \ref{tab:table_compare_final} summarizes the final results obtained for the three methods: Method 1 (the generalized model), Method 2 (different trainable parameters of each model but the same model architecture~\cite{tian2024attention}), and Method 3 (the proposed adaptive model). The comparison is based on various metrics, including MEE in three scenarios (S1, S2, and S3), the number of model parameters, adaptivity, and test time for S1. The test time is provided as a reference and may vary depending on the underlying hardware, software environment, and implementation details. All experiments are carried out in a Linux-based CPU-only environment with $64$\,GB of RAM. 

 \begin{table}[htbp!]
\caption{Final result comparison.}
\begin{center}
\begin{tabular}{|l|c|c|c|}
\hline
      & Method~1 & Method~2~\cite{tian2024attention} & Method~3 \\\hline
     S1, MEE (m)  & 0.47 & 0.99 & 0.40 \\\hline
     S2, MEE (m)  & 0.83 & 2.00 & 0.78 \\\hline
     S3, MEE (m)  & 0.86 & 1.01 & 0.77 \\\hline
     number of  & 303k & 227k + 227k & 87k + 126k \\
     parameters$^{\mathrm{a}}$ & & + 227k & + 126k \\\hline
     adaptive & + & - & + \\\hline
     test time, S1 (s)  & 4.40  & 1.83 & 1.48 \\\hline
     \multicolumn{4}{l}{$^{\mathrm{a}}$ Only one subset of the parameters is active at a given time} \\
     \multicolumn{4}{l}{for Method 2 and 3.} \\
  \end{tabular}
  \label{tab:table_compare_final}
  \end{center}
\end{table}

In \cite{tian2024attention}, the accuracy of Method~2 is given as $0.99$\,m, $2.00$\,m, and $1.01$\,m for S1, S2, and S3, respectively. In comparison, the proposed method shows improved accuracy, achieving MEE of $0.40$\,m, $0.78$\,m, and $0.77$\,m for S1, S2, and S3, respectively. Method~1, which is the generalized model, achieves an MEE of $0.47$\,m, $0.83$\,m, and $0.86$\,m for S1, S2, and S3, respectively, showing a slightly lower performance than the proposed method. The router system enables the activation of a single model at any moment, reducing the number of active parameters by $58\%$ to $71\%$ and the test time by $66\%$, compared to the generalized model. The test time of the router is negligible. 
Please note that these results are specific to the given scenarios, and adaptations may be needed for different deployment scenarios. For example, the deployment in new base station scenarios requires retraining of the model; however, this limitation is beyond the scope of this study.

\section{Conclusion}

In this work, we present an adaptive localization model that dynamically selects specialized sub-models based on the given input scenario. We compare our approach against two methods: (i) a manually switched model where each scenario is assigned to a specialized model, and (ii) a generalized model that attempts to handle all scenarios with a single model but suffers from degraded localization accuracy. The proposed model uses a routing mechanism to determine whether an input belongs to S1, S2, or S3, activating the corresponding sub-model accordingly. This approach enables the use of only a subset of the total parameters at any given time, reducing computational complexity compared to the generalized model, while achieving higher accuracy. 
To further improve efficiency, we introduce a pooling layer and remove layer normalization, reducing the model size and computational cost. The results highlight the potential of adaptive computation in radio-based localization, balancing accuracy and computational complexity across varying environmental conditions.

\section*{Acknowledgment}
The authors would like to thank Jens Gulin and Axel Berg from Lund University, as well as Wei Tang, Sangbu Yun, and the Lab for Efficient Application Processors (LEAPs) at the Department of Electrical Engineering and Computer Science, University of Michigan, Ann Arbor, for their valuable suggestions regarding our paper. 
This work is funded by the Swedish Research Council, Ericsson AB, and ELLIIT (Excellence Center at Linköping-Lund in Information Technology).
\bibliographystyle{ieeetr}
\bibliography{main}

\end{document}